\documentclass{appolb}
\usepackage{graphicx}
\usepackage{amsmath}
\usepackage{url}

\begin{document}
\title{Belle II prospects for CP-violation measurements
}
\author{Chiara La Licata for the Belle II Collaboration
\address{University and INFN, Trieste}
\\
}
\maketitle
\begin{abstract}
The Belle II collaboration operates a substantially upgraded Belle detector at the SuperKEKB energy-asymmetric $e^{+}$ $e^{-}$ collider. Belle II will start recording collisions at the $\Upsilon$(4S) energy in 2018, aiming to collect by 2025 50 ab$^{-1}$ of data, 50 times more than the Belle experiment. We report prospects for measuring quantities associated with charge-parity violation, with special emphasis on the Cabibbo-Kobayashi-Maskawa angle $\phi_3$($\gamma$) and observables in semileptonic $B$ meson decays associated with the quark-mixing matrix element $V_{ub}$.
\end{abstract}

\section{Introduction}
The Belle II detector~\cite{belle2} is currently at its final commissioning stage at the KEK laboratory in Tsukuba, Japan. The Belle II program is based on the success of the BaBar and Belle experiments, which discovered and established~\cite{physBfactories} violation of charge-parity symmetry (CP) in $B$ meson dynamics. The agreement with the expectations of the Standard Model of measurements related to CP violation (CPV) is limited to an accuracy of 10\%-20\%. Further studies of CP violation are therefore essential to over-constrain the dynamics in search for deviations from the Standard Model (SM). \\
Belle II is complementary to the LHC experiments. If LHC experiments will find evidence of non-SM physics, precision flavour inputs might likely be essential to characterise in more detail the novel dynamics. If no evidence for non-SM physics is found at LHC, the large samples collected by Belle II will provide a powerful indirect probe for new physics beyond the TeV scale. Belle II and LHCb are largely complementary. LHCb collects large samples of both $B^0_s$ and $B$ mesons and will dominate measurements based on final states with charged particles. Belle II is expected to lead in $B$ measurements of final states with neutrinos, or multiple photons. Belle II data taking will start at the beginning of 2019; the sample size equivalent to that of Belle will be reached in a few months, and in a couple of years it will reach 5 ab$^{-1}$. A data sample of 50 ab$^{-1}$ is expected to be collected within 7 years.

\section{SuperKEKB collider and Belle II detector}
The SuperKEKB accelerator is designed to provide an instantaneous luminosity of $8\times10^{35}$ cm$^{-2}$ s$^{-1}$, which exceeds by a factor of 40 that of KEKB. The target luminosity required a substantial upgrade of the accelerator complex including a completely novel collision approach, the “nano-beam” scheme~\cite{nano_beam}. The nano-scheme is based on the reduction of the beam vertical size at the interaction region, which provides a 20 fold increase in higher luminosity. The additional factor of 2 is obtained by doubling the beam currents.

\begin{figure}[ht!]
\centerline{%
\includegraphics[width=12.5cm]{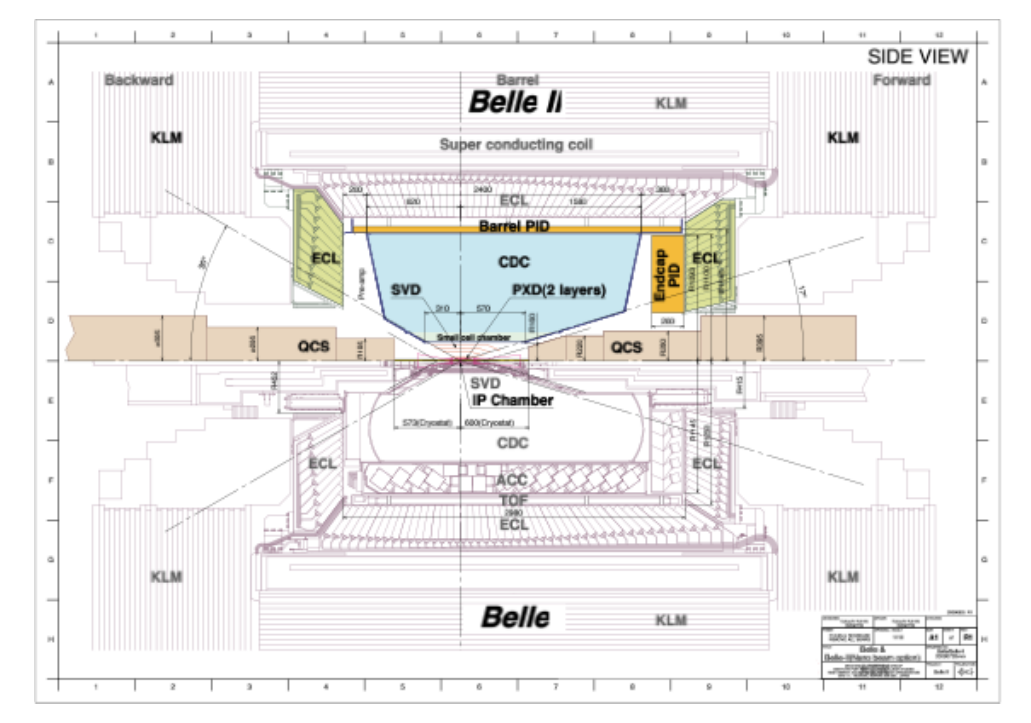}}
\caption{Belle vs Belle II detectors. The top half panel shows a side view of the Belle II detector, with new or upgraded components shown in colours; the bottom half panel shows the old Belle detector.}
\label{fig:belle2}
\end{figure}

As a consequence, the Belle II detector has to sustain 40 times higher event rates with backgrounds rates higher by a factor of 10 to 20 compared to Belle. Mitigating the effects of such higher backgrounds, which lead to an increase in the occupancy, pile up, fake hits and radiation damage, will be critical. All this required substantial modifications of the trigger, data acquisition and offline computing system with respect to Belle.
In addition many components of the Belle II detector have been upgraded or completely rebuilt (Fig.~\ref{fig:belle2}). Belle II features an entirely novel tracking, consisting of a pixel vertex detector, a silicon strip vertex detector, a central drift chamber, and new particle identification systems as well as a faster signal processing in the outer detectors.
The improved design choices translate into a factor of two expected improvement on the impact parameter resolution up to 20 $\mu m$ in both transverse and longitudinal components, a 30\% increase in $K^0_S$ efficiency, and a better $K/\pi$ separation, with $\pi$ misidentification rate decreased by a factor of 2.5.

\section{Decay-time-dependent CP violating observables in $B$ decays}
\begin{figure}[ht!]
\centerline{%
\includegraphics[width=12.5cm]{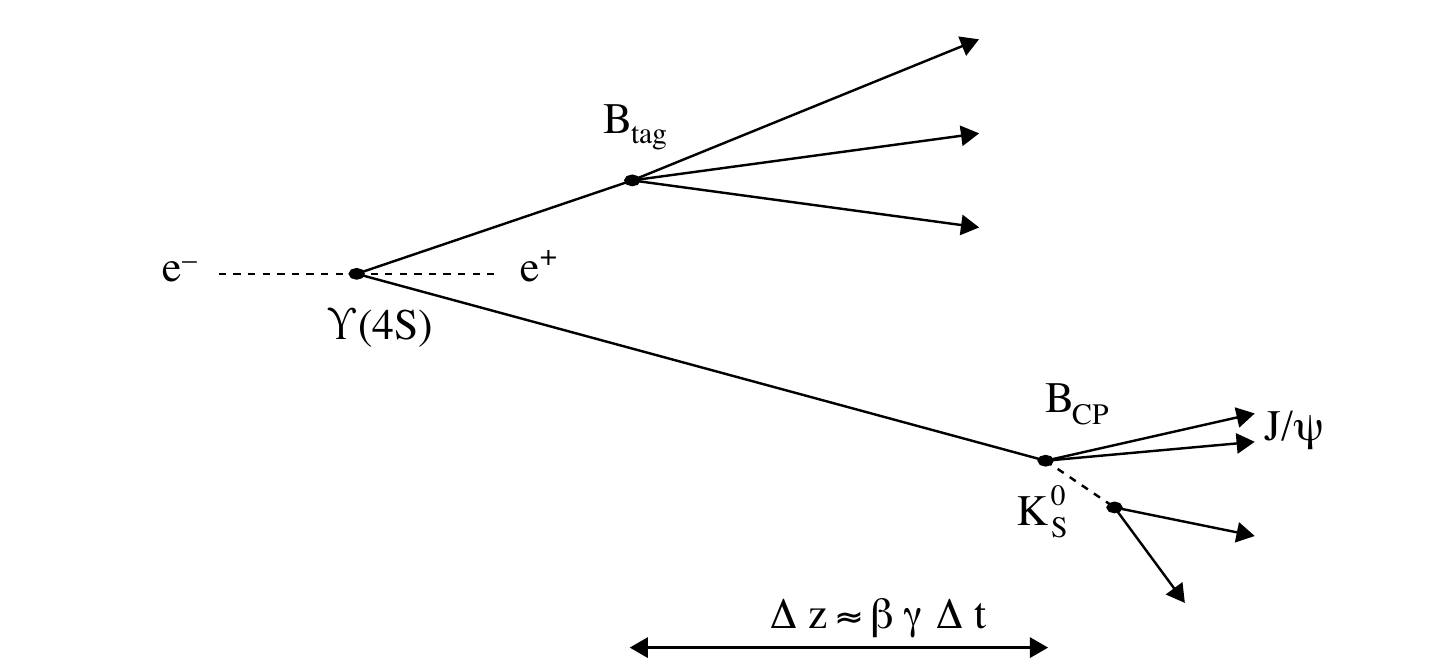}}
\caption{Decay sketch.}
\label{fig:timDep}
\end{figure}

CPV measurements represent the core of the Belle II program. This report focuses on those modes that are precisely measurable by Belle II, offering impact unique or competitive with LHCb.
Precision measurements of the Cabibbo-Kobayashi-Maskawa (CKM) phases $\phi_1$\footnote{Different notations are used for CKM phases. In this report we use $\phi_1$ ($\equiv \beta$), $\phi_2$ ($\equiv \alpha$) and $\phi_3$ ($\equiv \gamma$).} ($\equiv$ arg[-V$_{cd}$V$_{cb}^{\ast}$/(V$_{td}$V$_{tb}^{\ast}$)] where $V_{ij}$ are the couplings of quark mixing transitions from an up-type quark i = $u$, $c$, $t$ to a down-type quark j = $d$, $s$, $b$) and $\phi_2$ ($\equiv$ arg[-V$_{td}$V$_{tb}^{\ast}$/(V$_{ud}$V$_{ub}^{\ast}$)]) are crucial inputs to the CKM unitarity triangle fits.
A general strategy for extracting the $\phi_1$ and $\phi_2$ angles is based on decay-time-dependent CP asymmetry measurements. Asymmetric $e^{+}$ $e^{-}$ $B$ factories are ideal environments to perform such measurements as neutral $B$ mesons are produced through the transition $e^{+} e^{-} \rightarrow \Upsilon \rightarrow B^{0} \bar{B^{0}}$ in an entangled state. Both $B$ mesons evolve in this coherent state, where there are exactly one $B^{0}$ and one $\bar{B^{0}}$, until either meson decays. At that point the other $B$ meson continue to propagate and oscillate between a $B^{0}$ and $\overline{B^{0}}$ state until its own decay. The identification, through its decay, at any given time $t$, of one of the $B$ mesons as a $B^0$ (or $\bar{B^{0}}$) offers information on the flavor of the other $B$ meson at the same time (Fig.~\ref{fig:timDep}). CP violation is studied using decay-time-dependent asymmetries of decay rates into CP eigenstates,
\begin{equation}
    a_{f_{\rm CP}(\Delta t)} = \frac{\Gamma[\bar{B}(\Delta t)]-\Gamma[B(\Delta t)]}{\Gamma[\bar{B}(\Delta t)]+\Gamma[B(\Delta t)} = C cos(\Delta m \Delta t) - S sin(\Delta m \Delta t),
\end{equation}
where $\Delta t$ is the interval between the time $t_0$ at which the flavor of the $B$ meson is observed and the time where the meson decays into the CP eigenstate and $\Delta m$ is the mass difference. The decay time is inferred from the decay length through $\Delta t = \Delta z/\beta \gamma c$ (where $z$ is the boost direction, $\gamma$ is the Lorentz factor and $c$ is the speed of light.).  The $C$ coefficient is related to "direct" CP violation while $S$ is related to the "mixing induced" CP violation and it corresponds to a particular combinations ($\phi_i^{\rm eff}$) of the angles of the Unitarity Triangle through $S=sin(\phi_i^{\rm eff})$.

Two key aspects need be considered, the $\Delta t$ resolution related to vertex fitter performance and the flavour tagger. The most important contribution to the $\Delta t$ resolution comes from the vertex fit of the tag side. The reduction of the boost and the expected higher beam-related backgrounds with respect to Belle make this task expecially challenging. However, the pixel detector together with the new vertex reconstruction algorithms are expected to provide a net improvement over Belle. That translates to a $\Delta t$ with an expected resolution of 0.77 ps. \\
The tagging efficiency, defined as the fraction of events for which a flavor tag is assigned, is also critical. Simulation predicts a tagging efficiency of 35.8\% for Belle II which would represent a 20\% improvement over Belle performance.

\section{Sensitivity on $\phi_1$}
The parameter $\phi_1$ (=21.4$^\circ$ $\pm$ 0.8$^\circ$) is currently the best known Unitarity Triangle (UT) angle.
The most effective way for measuring $\phi_1$ involves $B^{0}$ decays to CP eigenstates governed by $b \rightarrow c\bar{c}s$ quark transitions. The "golden mode" for these transitions is $B \rightarrow J/\psi K^{0}_S$ as the detection of the final state and the theoretical interpretation are especially straightforward. The $J/\psi$ decays to lepton pairs and the $K^0_S$ is sufficiently long-lived to decay into pairs of opposite charged pions at a secondary vertex largely displaced from the interaction region.
The straightforward signature together with a relatively large branching fraction translate into an expected large signal yield.  Another net advantage is that the contribution of higher-order amplitudes that are difficult to calculate, is expected to be less than $1\%$. Their contributions, however, may become more important with the full Belle II sample and should be taken into account for the final $\phi_1$ determination.

The current precision on the world-average values for $S$ from the $J/\psi K_S^{0}$ final state is dominated by the statistical uncertainty~\cite{CKMfitter}. The most important uncertainty for $C$ comes from systematic sources.
For an extrapolation of the statistical precision reachable by Belle II, we assume the same $B\bar{B}$ vertex separation capability of Belle and scale the uncertainty according to the inverse square root of the integrated luminosity.
Many sources of systematic uncertainties are expected in the wrong-tag fractions, possible fit biases, signal fractions, the background $\Delta t$ distribution, $B$ meson lifetime and mass difference. All depend on the simulated sample sizes and are therefore expected to be reducible. Two remaining irreducible uncertainties need a dedicated study, the tag side interference and the uncertainty due to the vertex reconstruction algorithm.
Table~\ref{tab_expPrecisionPhi_1} shows the expected results in 50 ab$^{-1}$\cite{b2tip}. They are expected to be dominated by systematic uncertainties. Three scenarios are considered, 'no improvement', where no improvements are assumed over the Belle irreducible systematic uncertainties; 'vertex improvement', where an improvement of 50\% is assumed for the systematic uncertainty due to the vertex positions; and 'leptonic categories', where the analysis uses only leptonic categories for flavour tagging.
A precision better than 1\% is expected when combining all the $b \rightarrow c\bar{c}s$ modes with the full Belle II data set.

\begin{table}[]
    \centering
    \begin{tabular}{cccc}
    \hline \hline
     $J/\psi K_S^{0}$ & 'no improvement' & 'vertex improvement' & 'leptonic categories' \\ \hline
     S (50 ab$^{-1}$) & & & \\
    Stat. & 0.0035 & 0.0035 &  0.0060 \\
    Syst. reducible & 0.0012 & 0.0012 &  0.0012 \\
    Syst. irreducible & 0.0082 & 0.0044 &  0.0040 \\
    \hline
     C (50 ab$^{-1}$) & & & \\
    Stat. & 0.0025 & 0.0025 &  0.0043 \\
    Syst. reducible & 0.0007 & 0.0007 &  0.0007 \\
    Syst. irreducible & +0.043,-0.022 & +0.042,-0.011 &  0.011 \\
    \hline \hline
    \end{tabular}
    \caption{Belle II expected sensitivity on the CP parameters for the $J/\psi K_S^{0}$ decay mode.}
    \label{tab_expPrecisionPhi_1}
\end{table}

The decay-time-dependent CP asymmetry parameters $S$ can be also measured in charmless decays to CP eigenstates governed by the $b \rightarrow sq\bar{q}$ quark transition. These decays are particularly sensitive to non-SM physics because any unobserved heavy particle with compatible quantum numbers could contribute an additional penguin loop and alter the value of the observed weak phase. A significant deviation of the measured value of $S$ in one or more charmless decays from the value from tree-dominated processes could indicate non-SM physics. Belle II performed a sensitivity study for the $B^{0}\rightarrow \phi K^{0}$ and $B^{0}\rightarrow \eta^{'} K^{0}$ modes.
Both the BaBar and Belle experiments extracted the $B^{0}\rightarrow \phi K^{0}$ CP asymmetry parameters from the time-dependent analysis of the $K^{+}K^{-}K^{0}$ final state \cite{b_phi_K0_babar,b_phi_K0_belle}.
The Belle II decay-time-dependent study considers four benchmark final states in a quasi-two body approach, $\phi(K^{+}K^{-}) K^{0}_{s}(\pi^{+}\pi^{-})$, $\phi(K^{+}K^{-}) K^{0}_{s}(\pi^{0}\pi^{0})$, $\phi(\pi^{+}\pi^{-}\pi{0}) K^{0}_{s}(\pi^{+}\pi^{-})$, and $\phi(K^{+}K^{-}) K_{L}$.
Table~\ref{tab_phi_K0} lists the expected sensitivity estimated for an integrated luminosity of 5 ab$^{-1}$.

\begin{table}[h!]
    \centering
    \begin{tabular}{ccccc}
    \hline \hline
         Channel & $\epsilon_{reco}$ & Yield & $\sigma(S_{\phi K^{0}})$ & $\sigma(A_{\phi K^{0}})$ \\ \hline
         $\phi(K^{+}K^{-})$ $K^{0}_{S}(\pi^{+}\pi^{-})$ & 35\% & 2280 & 0.078 & 0.055 \\
         $\phi(K^{+}K^{-})$ $K^{0}_{S}(\pi^{0}\pi^{0})$ & 25\% & 765 & 0.132 & 0.096 \\
         $\phi(\pi^{+}\pi^{-}\pi^{0})$ $K^{0}_{S}(\pi^{+}\pi^{-})$ & 28\% & 545 & 0.151 & 0.113 \\
         $K^{0}_S$ modes combination & & & 0.060 & 0.044 \\
         $K^{0}_S+K^{0}_L$ modes combination & & & 0.048 & 0.035 \\ \hline \hline

    \end{tabular}
    \caption{Sensitivity estimates for $S_{\phi K_{0}}$ and $A_{\phi K_{0}}$ parameters, the expected yield and the reconstruction efficiency considering an integrated luminosity of 5 ab$^{-1}$ in the $B^{0}\rightarrow \phi K^{0}$ decay mode.}
    \label{tab_phi_K0}
\end{table}
Another promising decay in Belle II is $B^{0}\rightarrow \eta^{'} K^{0}$. Previous BaBar and Belle results are dominated by statistical uncertainties $S_{\eta^{'} K^{0}}$ = $+$0.57 $\pm$ 0.08 $\pm$ 0.02 (BaBar) and $S_{\eta^{'} K^{0}}$ = $+$0.68 $\pm$ 0.07 $\pm$ 0.03 (Belle)~\cite{etaPK_belle}.
Belle II performed a sensitivity study considering the benchmark decay channels $\eta^{'}(\eta(\gamma \gamma)\pi^{+}\pi^{-})K^{0}_S(\pi^{+}\pi^{-})$ and $\eta^{'}(\eta(\pi^{+}\pi^{-}\pi^{0})\pi^{+}\pi^{-})K^{0}_S(\pi^{+}\pi^{-})$. The expected uncertainties for $S_{\eta^{'}K^{0}_{S}}$ and $A_{\eta^{'}K^{0}_{S}}$ with a 5 ab$^{-1}$ data set are listed in Tab.~\ref{tab_etaPrimeK}

\begin{table}[h!]
    \centering
    \begin{tabular}{ccccc}
    \hline \hline
         Channel & $\sigma(S_{\eta^{'}K^{0}_{S}})$ & $\sigma(A_{\eta^{'}K_{S}})$ \\ \hline

         $\eta^{'}(\eta(\gamma \gamma)\pi^{+}\pi^{-})$ $K^0_{S}(\pi^{+}\pi^{-})$ & 0.06 & 0.04 \\

         $\eta^{'}(\eta(\pi^{+}\pi^{-}\pi^{0})\pi^{+}\pi^{-})K_S^{0}(\pi^{+}\pi^{-})$ & 0.11 & 0.08 \\

         $K^{0}_{S}$ modes & 0.028 & 0.021 \\

         $K^{0}_{S}+K^{0}_{L}$ modes & 0.027 & 0.020 \\

         \hline \hline

    \end{tabular}
    \caption{Sensitivity estimates for $S_{\eta^{'}K_{S}}$ and $A_{\eta^{'}K_{S}}$ parameters for an integrated luminosity of 5 ab$^{-1}$ in the $B^{0}\rightarrow \eta^{'} K^{0}$ decay.}
    \label{tab_etaPrimeK}
\end{table}

\section{Sensitivity on $\phi_2$}
\begin{figure}[ht!]
\centerline{%
\includegraphics[width=12.5cm]{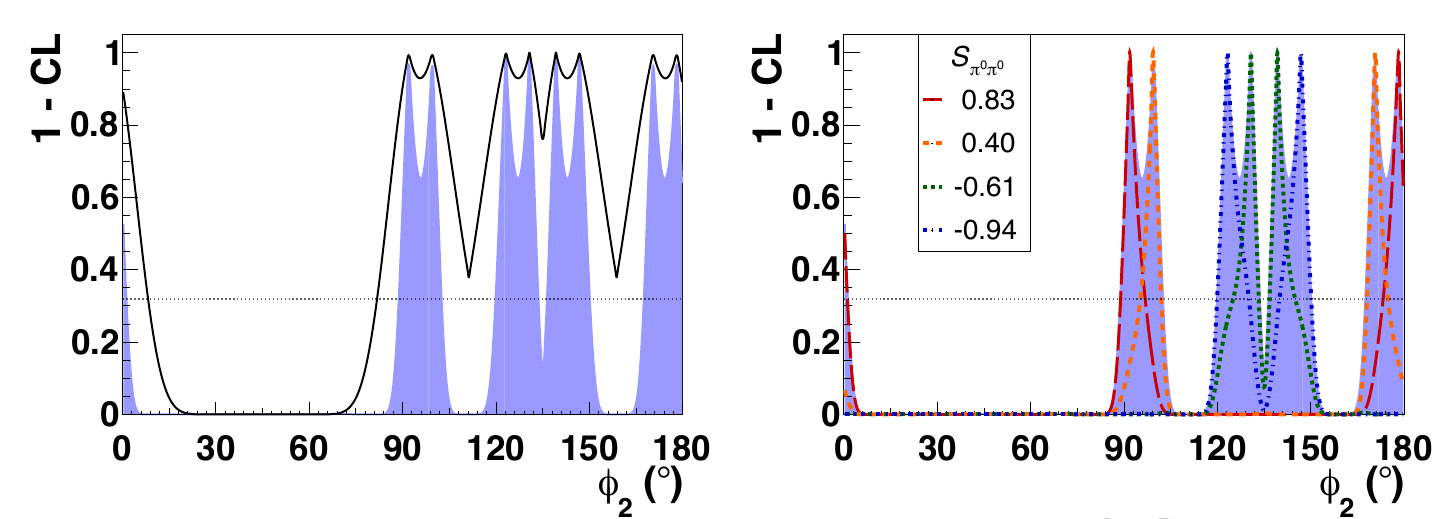}}
\caption{P-value vs $\phi_2$ for the isospin analysis of the $B \rightarrow \pi \pi$ system. The black solid line (left) shows the result of the scan using data from Belle measurements. The blue shaded area in both plots shows the projection for Belle II. Results of the scan adding the $S_{\pi^0 \pi^0}$ constraint (right); each line shows the result for a different $S_{\pi^0 \pi^0}$ value. The dotted horizontal lines correspond to one standard deviation.}
\label{fig_phi_2_plots}
\end{figure}

The $\phi_2$ angle is accessible through the measurement of the decay-time-dependent CP asymmetry of $B$ meson decays into $\pi \pi$ and $\rho \rho$ final states. These final states are also accessible by higher-order weak transitions including penguin amplitudes. The presence of penguin contributions with weak phases differing from the leading-order phase results in an additional phase $\Delta\phi_2$ that biases the parameter $S$.
The most effective way to extract $\Delta\phi_2$ is to apply an isospin analysis to $B \rightarrow \pi \pi$ and $B \rightarrow \rho \rho$ decays.
An essential input for the isospin analysis of $B \rightarrow \pi \pi$ decays is the value of $S$ parameter for $B^{0} \rightarrow \pi^0 \pi^0$ decays, which has not yet been measured.
In general, an eightfold ambiguity exists in the $\phi_2$ solutions obtained from the isospin analysis without $S_{\pi^{0}\pi^{0}}$ information. This reduces to a two-fold ambiguity once $S_{\pi^{0}\pi^{0}}$ is folded in. Belle II has performed $B^{0} \rightarrow \pi^0 \pi^0$ sensitivity studies exploiting the following final states:
\begin{itemize}
    \item $B^{0} \rightarrow \pi^0(\rightarrow \gamma \gamma) \pi^0(\rightarrow \gamma \gamma)$
    \item $B^{0} \rightarrow \pi^0(\rightarrow e^{+}e^{-}\gamma) \pi^0(\rightarrow \gamma \gamma)$
    \item $B^{0} \rightarrow \pi^0(\rightarrow \gamma(\rightarrow e^{+}e^{-})\gamma) \pi^0(\rightarrow \gamma \gamma)$.
\end{itemize}
The last two decay modes are particularly important. Only decays where one of the two $\pi^{0}$ decays to three bodies ($\pi^{0}\rightarrow e^{+}e^{-}\gamma$) or when one photon from the neutral pions converts in, or before, the vertex detector provide a precise measurement of the $B^{0}$ vertex.
The isospin analysis will also benefit from reduced uncertainties on currently used input observables. Fig.~\ref{fig_phi_2_plots} left shows the expected improvement in the determination of $\phi_2$ when all the isospin analysis inputs are included with a reduced uncertainty considering an integrated luminosity of 50 ab$^{-1}$. The right panel of Fig.~\ref{fig_phi_2_plots} shows the results when the $S_{\pi^0 \pi^0}$ input is included. \\

\begin{figure}[ht!]
\centerline{%
\includegraphics[width=6cm]{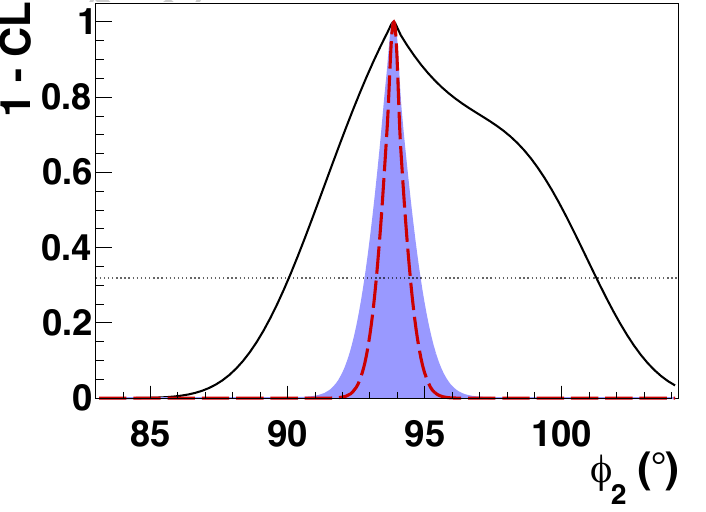}}
\caption{P-value vs $\phi_2$ for the isospin analysis of the $B^{0} \rightarrow \pi \pi$ and the $B^{0} \rightarrow \rho \rho$ decay modes. The black solid line shows the results of the scans using data from measurements at the current precision. The blue shaded area shows Belle II projections. The red dashed line shows the results of the scans adding the $S_{\pi^{0}\pi^{0}}$ and $S_{\rho^{0}\rho^{0}}$ constraints.}
\label{fig_tot_phi2}
\end{figure}

The $B \rightarrow \rho \rho$ mode will also be exploited in the isospin analysis. Fig.~\ref{fig_tot_phi2} shows the expected improvement in the determination of $\phi_2$ by combining the isospin analyses of $B \rightarrow \pi \pi$ and $B \rightarrow \rho \rho$. We expect a 1$\sigma$ statistical uncertainty of about 0.6$^\circ$.

\section{Sensitivity on $\phi_3$}
Belle II has performed a dedicated study of its potential to determine the unitarity triangle angle $\phi_3$ ($\equiv$arg[-$V_{ud}V^{\ast}_{ub}$/$V_{cd}V^{\ast}_{cb}$])
using the interference between the $b\rightarrow c\bar{u}s$ and $b\rightarrow u \bar{c}s$ tree amplitudes in the charged $B$ decay to final states $B\rightarrow DK$.
The key feature is that such decays arise solely from the interference of tree level diagrams of differing weak and strong phases. No $B$ mixing, nor penguin amplitudes, are involved. That allows a theoretically robust extraction of $\phi_3$ with theoretical uncertainty much smaller than 1\%. However, knowledge of $\phi_3$ is still limited by the small branching fractions of the processes involved.
Currently, the most precise determination of $\phi_3$ is reported by the LHCb experiment ($\phi_3$=72.2$^\circ$\textsuperscript{+6.8$^\circ$}\textsubscript{-7.3$^\circ$}~\cite{phi3_lhcb}).
Several techniques allow for extracting $\phi_3$, depending on $D$ decay considered. Belle II investigated the sensitivity using the Dalitz plot analysis of the "golden mode" $B^{\pm} \rightarrow [K^{0}_{S}\pi^{+}\pi^{-}]_{D}K^{\pm}$. The relatively large branching fraction and effective Belle II $K^0_S$ reconstruction efficiency makes this mode particularly promising. Important inputs to this analysis are the strong phases between the $D$ and $\bar{D}$ decays restricted to each bin of the Dalitz plot. These are provided by the charm factories. The systematic biases related to the uncertainties on such inputs are expected to be significant for Belle II. Based on simulated $B^{\pm} \rightarrow D(\rightarrow K^{0}_{S}\pi\pi)K^{\pm}$ decays in 50 ab$^{-1}$ and assuming 10 fb$^{-1}$ collected at the $\Upsilon(3770)$ resonance at BESIII~\cite{besIII}, the $\phi_3$ precision is expected to reach $3^{\circ}$. 
With the inclusion of results obtained from the analysis of $D$ mesons decaying to CP-eigenstates, Cabibbo-favoured and doubly Cabibbo suppressed final states, such figure is expected to reduce to $1.6^{\circ}$.
We expect a close competition in the Belle II and LHCb precision. The combination of results from both experiments will resonably push the knowledge of $\phi_3$ to better than 1\%.

\section{Sensitivity on $V_{ub}$}
Semileptonic $B$ decays proceed via first-order weak interactions. These are mediated by the $W$ boson and are expected to be free of non-SM contributions. Therefore such decays play a key role in the determination of the magnitude of the CKM matrix element $V_{ub}$. A precise measurement of the length of the side of the unitarity triangle opposite to $\phi_1$ is particularly important as a consistency check of the SM picture. The length of this side is determined to good approximation by the ratio of the magnitudes of two CKM matrix elements, $|V_{ub}|/|V_{cb}|$.
The $|V_{ub}|$ parameter is measured through two approches, one based on the study of exclusive decays and the other that involves inclusive decays.
Currently the values of the inclusive and exclusive determinations of $|V_{ub}|$ are inconsistent at a level of 4$\sigma$,
\begin{align*}
|V^{\text{incl}}_{ub}| = (4.52 \pm 0.15 (\text{exp})^{+0.11}_{-0.14}(\text{theo}))\times 10^{-3}
\end{align*}
\begin{align*}
|V^{\text{escl}}_{ub}| = (3.67 \pm 0.09 (\text{exp})\pm 0.12(\text{theo}))\times 10^{-3}.
\end{align*}

\subsection{Exclusive measurement}
The study of the exclusive semileptonic decay $B \rightarrow \pi l \nu$ offers access to $|V_{ub}|$. Its decay rate, in the limit of $m_\ell \rightarrow 0$, which is a good approximation for $\ell$ = $e$, $\mu$, assumes the following form:

\begin{align*}
\frac{d\Gamma}{dq^{2}} = \frac{G^{2}_{F}|V_{ub}|^2}{24 \pi^{3}}p^3_{\pi}|f_+^{B\pi}(q^{2})|^2,
\end{align*}
where $G_F$ is the Fermi constant, $f_+^{B\pi}(q^{2})$ is the form factor, q is the momentum transfer and $p_{\pi}$ is the pion 3-momentum in the B rest frame.
Hence, precise experimental measurements of the $B\rightarrow \pi l \nu$ branching fraction, together with reliable theoretical calculations of the form factor $f_+^{B\pi}(q^{2})$, allow for a theoretically robust $|V_{ub}|$ determination. Currently the $|V_{ub}|$ measurement is limited by the precision of the form factor calculated by lattice QCD and light-cone sum rules~\cite{exclusive_Vub}.
Two techniques for the measurement of the differential branching fraction exist, using high-purity hadronic $B$ tag reconstruction \cite{tagg_Vub_exc}, and low-purity untagged measurements \cite{untagg_Vub_exc}.
In the tagged measurement, the $B$ companion is fully reconstructed in one of the many hadronic decay modes. Then the rest of the event is required to be consistent with the signature of the signal decay (only two additional opposite charged particles, one being consistent with the pion and one with the lepton hypothesis). In untagged measurements, the signal $B$ candidate is reconstructed by selecting quality pion and lepton candidates originating from the same space-point.
As the Belle II detector covers hermetically a large portion of the full solid angle, we assume that all remaining tracks and clusters of the rest of the event came from the companion $B$ meson, $B_{comp}$. The $B_{comp}$ is then reconstructed by summing the four-momenta of the hypothetical particles from the remaining tracks and clusters.
The precise measurement of the $B_{comp}$ momentum in the tagged analysis implies an improved determination of $q^{2}$ compared to the untagged measurement. The simulated reconstruction efficiency is 0.55\%, which is considerably above the reconstruction efficiency (0.3\%) of the tagged measurement reported by Belle.
The expected precision of the exclusive $|V_{ub}|$ determination from the untagged (tagged) analysis is 1.7\% (1.3\%) on the 50 ab$^{-1}$ data set assuming the expected factor of 5 improvement in the progress in lattice quantum chromodynamics calculations over the next decade.\\

\subsection{Inclusive measurement}
$|V_{ub}|$ can also be measured from the inclusive rate, by summing over all hadronic final states subject to appropriate kinematic constraints. The Belle inclusive analysis shows significant systematic uncertainties related to modelling of the charmless semileptonic decay composition. These uncertainties might be reduced with more data through a better characterisation of $B\rightarrow X_{u}l\nu$ in hadron-tagged samples.
The background from $B\rightarrow X_{c}l\nu$ decays, which is 50 times larger than the signal, makes the extraction of $|V_{ub}|$ from the inclusive measurement very challenging. For this measurement, kinematic selection criteria have to be applied to suppress $B\rightarrow X_{c}l\nu$ decays, and $|V_{ub}|$ is extracted from the differential $B$ $\rightarrow X_{u} l \nu$ rate in various phase space regions. The full data sample at Belle II is expected to allow for more a precise characterisation of high mass $B\rightarrow X_c l \nu$ modes, with more accurate background subtraction in low purity, high mass regions.
Moreover the Belle II large data set is expected to allow precise, unfolded measurements of the kinematic spectra in $B\rightarrow X_u l \nu$ decays. That will give important information about shape functions, weak annihilation and signal modelling, thus testing their validity.
A projection to the first 5 ab$^{-1}$ of Belle II data with two single differential spectra in the invariant mass of the hadronic part of the decay, and the lepton energy of $B\rightarrow X_u l \nu$, shows a factor 2 of improvement over current results.

\section{Summary}
Belle II is in its final stage of commissioning toward physics data taking, which will start in early 2019. The plan is to reach five-fold the Belle sample size in a couple of years and a further factor of ten is expected to be collected within the subsequent five years.
The large data set, together with substantial improvements in the detector, are expected to allow for significant improvements in many measurement of CKM parameters. The existing constraints, in fact, do not exclude the possibility of sizeable contributions from non-SM processes in various flavour physics observables that can be tested in the next few years.
The parameter sin($2\phi_1$) is expected to reach 1\% a precision, remaining the most precise input associated with the Unitarity Triangle.
The parameter sin($2\phi_2$) will benefit from the inclusion of additional and improved isospin analysis inputs that will reduce ambiguities in the $\phi_2$ solution.
The $\phi_3$ precision is expected to reach $1.6^\circ$. $|V_{ub}|$ will be measured with a precision up to 1.3\%(3\%) in exclusive (inclusive) semileptonic measurements. Thanks to these and many other results, Belle II will drive a significant advance in our knowledge the quark flavour dynamics.

\end{document}